\documentclass[12pt]{article}

\usepackage{sbc-template}

\usepackage{subfigure}
\usepackage{makecell}

\usepackage{soul}
\usepackage{color}
\usepackage{subfigure}

\usepackage{graphicx,url}

\usepackage[utf8]{inputenc}

\title{On the Performance of Cyber-Biomedical Features for Intrusion Detection in Healthcare 5.0}

\author{Pedro H. Lui\inst{1}, Lucas P. Siqueira\inst{1}, Juliano F. Kazienko\inst{1}, Vagner E. Quincozes\inst{2},\\ Silvio E. Quincozes\inst{3}, and Daniel Welfer\inst{1}}
\address{Universidade Federal de Santa Maria (UFSM), Santa Maria -- RS, Brasil
\vspace{-2mm}
\nextinstitute
Universidade Federal Fluminense (UFF), Niterói -- RJ, Brasil
\vspace{-2mm}
\nextinstitute
Universidade Federal do Pampa (UNIPAMPA), Alegrete -- RS, Brasil \vspace{-1mm}
\email{\{pedro.lui, kazienko\}@redes.ufsm.br, lucas.pittella@acad.ufsm.br} \vspace{-2mm}
\email{vequincozes@id.uff.br, silvioquincozes@unipampa.edu.br} \vspace{-2mm}
\email{daniel.welfer@ufsm.br}
}

\begin{document} 

\maketitle

\begin{abstract}
Healthcare 5.0 integrates Artificial Intelligence (AI), the Internet of Things (IoT), real-time monitoring, and human-centered design toward personalized medicine and predictive diagnostics. However, the increasing reliance on interconnected medical technologies exposes them to cyber threats. Meanwhile, current AI-driven cybersecurity models often neglect biomedical data, limiting their effectiveness and interpretability. This study addresses this gap by applying eXplainable AI (XAI) to a Healthcare 5.0 dataset that integrates network traffic and biomedical sensor data. Classification outputs indicate that XGBoost achieved 99\% F1-score for benign and data alteration, and 81\% for spoofing. Explainability findings reveal that network data play a dominant role in intrusion detection whereas biomedical features contributed to spoofing detection, with temperature reaching a Shapley values magnitude of~0.37.




\end{abstract}

     

\section{Introduction} \label{sec:intro}

The advent of Healthcare 5.0 marks a transformation in medical innovation. Building on Healthcare 4.0's technological integration, it shifts the focus to patient-centered treatment and care, integrating Artificial Intelligence (AI), the Internet of Things (IoT), and human-centered design to enhance personalized medicine, predictive diagnostics, and real-time monitoring.  These technologies are also reshaping cybersecurity in healthcare, offering new information (features) and generating opportunities to enhance data-driven security mechanisms~\cite{Gadekallu}.

The exponential growth of the Digital Health market, projected to reach \$258.25 billion by 2029~\cite{statistadigitalhealth2025}, underscores the critical need to prioritize cybersecurity in the Internet of Medical Things (IoMT). As connected biosensors, telemedicine platforms, and digital treatment tools become integral to healthcare delivery, their reliance on sensitive patient data and connectivity exposes vulnerabilities to cyberattacks. A single breach could compromise patient safety, disrupt critical care, and erode trust in rapidly expanding markets. As healthcare becomes increasing reliant on digital solutions, accelerated by post-pandemic adoption, proactive research into IoMT cybersecurity is essential to safeguard data integrity, ensure regulatory compliance, and sustain the sector’s growth. Addressing these risks now will protect both technological innovation and human lives.

In cybersecurity, AI is instrumental in detecting threats through network traffic analysis, with Intrusion Detection Systems (IDS) based on Machine Learning (ML) algorithms to identify malicious patterns and strengthen network defenses. However, a significant limitation arises in the era of Healthcare 5.0: many AI-driven cybersecurity solutions are tailored for non-healthcare domains~\cite{HadyIDS}, and rely on datasets lacking critical biomedical data from wearable devices and medical sensors. This absence restricts their usefulness in healthcare, diminishes transparency, and complicates incident response efforts. To overcome this limitation, recent initiatives have focused on creating integrated datasets that combine network traffic with biomedical sensor data (\textit{i.e., Cyber-Biomedical features}), alongside the development of Explainable AI (XAI) frameworks to improve interpretability and decision-making \cite{ali2023explainable}\cite{alani2023xmednn}\cite{aljuhani2023intelligent}\cite{sohail2024explainable}. However, such initiatives emphasize network traffic features while overlooking the role of XAI in assessing biomedical data contributions to intrusion detection. This gap limits IDS effectiveness in detecting adversarial manipulations of biometric signals---such as falsified heart rate or oxygen levels---potentially compromising patient safety.




In this work, we evaluate the predictive contributions of biomedical and network features in IDSs for IoT-driven Healthcare 5.0 environments. For analysis, SHAP (SHapley Additive exPlanations)\footnote{SHAP Tool. Available at: \url{https://shap.readthedocs.io/en/latest/}} was used as a XAI tool to analyze feature prediction relevance from the WUSTL-EHMS-2020 dataset \cite{HadyIDS}, which joins network and biomedical features. Our results indicate that network data plays a primary role in intrusion detection, while biomedical data are highly relevant for the detection of cyber-physical attacks that manipulate biometric signals. These findings expose a major shortcoming in current medical cybersecurity research: the failure to integrate biomedical data into IDS models, despite the growing interconnectivity of Cyber-Physical Systems (CPS) in the healthcare domain. By providing a transparent assessment of feature importance, this study reinforces the need to merge real-time network analytics with biomedical insights, strengthening Healthcare 5.0 cybersecurity to patient safety and data integrity.



The paper is structured as follows: Section~\ref{sec:back} introduces key concepts, while Section~\ref{sec:tr} reviews related work in Healthcare 5.0 and identifies research gaps. Section~\ref{sec:material} details the methodology, covering experimental setups, AI models, and evaluation metrics. Section~\ref{sec:results} presents the results, including SHAP-based feature analysis and interpretations. Finally, Section~\ref{sec:conclusao} concludes the study and discusses future research directions.

\section{Background} \label{sec:back}


This section introduces key concepts and technologies, including Healthcare 5.0, the role of machine learning and XAI in decision-making and transparency, and the importance of IDS for healthcare cybersecurity. 

\subsection{Healthcare 5.0} \label{subsec:healthcare50}

Healthcare 5.0 represents the next stage in healthcare evolution, characterized by personalized, proactive, and patient-centered treatment enabled by advanced technologies such as smart wearables, ML, and the IoMT. Devices like biosensors and fitness trackers enhance clinical decision-making and patient self-management by enabling real-time data analysis, remote care, and continuous health monitoring. By integrating ML and IoMT, Healthcare 5.0 advances predictive analytics, early disease detection, and tailored treatments, significantly improving healthcare efficiency and patient outcomes~\cite{tandel2024intelligent}.

However, as connected medical devices become increasingly prevalent, cybersecurity emerges as a critical concern. The interconnectivity of smart healthcare systems exposes patient data and medical infrastructure to cyber threats, necessitating robust security mechanisms. AI-driven cybersecurity solutions play a pivotal role in mitigating these risks by detecting anomalies, identifying potential intrusions, and ensuring the reliability of connected healthcare devices~\cite{khan2024explainable}. Ensuring the security and integrity of medical data is essential for realizing the full potential of Healthcare 5.0, reinforcing the need for solutions that balance innovation with privacy and system resilience.

\subsection{Machine Learning and Explainable AI} \label{subsec:mlxai}

ML plays a fundamental role in AI, enabling automation across a wide range of tasks. 
However, most ML models operate as black boxes, making their decision-making processes
opaque and difficult to interpret~\cite{ali2023explainable}. This lack of transparency has led to the rise of XAI, which seeks to enhance model interpretability while maintaining high accuracy. In healthcare, where decision-making must be transparent and justifiable~\cite{dave2020explainable}, this challenge is particularly critical. The inability to explain AI-driven predictions reduces clinical trust, as the risks associated with unverified or uninterpretable AI recommendations may outweigh benefits in accuracy, speed, and efficiency. Thus, XAI plays a crucial role in increasing trust, improving accountability, and ensuring the safe integration of AI in medical applications.

\subsection{Intrusion Detection Systems} \label{subsec:ids}

IDSs play a crucial role in cybersecurity by continuously monitoring network activity and identifying anomalous behavior~\cite{JaveedIDSXAI}. AI-driven IDSs have gained prominence due to their ability to rapidly analyze vast amounts of data, uncovering complex attack patterns that traditional security methods might miss. These systems are particularly valuable in high-stakes environments such as the IoMT and critical infrastructure, where security breaches can have severe operational and safety implications. In healthcare, IDSs are essential for detecting cyber threats targeting connected medical devices, hospital networks, and remote monitoring platforms, where a single vulnerability could jeopardize patient safety and data integrity~\cite{alani2023explainable}. These IDSs are typically trained on datasets composed of features and samples, which must be representative to ensure accurate detection of both normal and malicious network behavior. As IoMT adoption continues to expand, ensuring the robustness of IDS solutions becomes increasingly urgent to protect interconnected healthcare systems from evolving cyber threats.

\section{Literature Review} \label{sec:tr}

In recent years, an expanding body of research has investigated the integration of Healthcare 5.0 principles with cybersecurity frameworks. This section reviews studies in this domain. Additionally, a comparative analysis is presented in Table~\ref{tab:relatedworks}, which evaluates the approaches and limitations of these works in relation to the current study.

\begin{table}[ht]
\centering
\caption{Comparison to the Related Works.}
\label{tab:relatedworks}
\small
\begin{tabular}{|c|c|c|c|c|} 
    \hline
    \textbf{Reference} & 
    \makecell{\textbf{Biomedical Feature}  \\ \textbf{Impact Analysis on IDS}} & 
    \makecell{\textbf{Biomedical Feature} \\ \textbf{Merged with IDS}} &
    \makecell{\textbf{EHMS} \\ \textbf{Dataset}} &
    \textbf{XAI} \\    
    \hline
    \cite{alani2023xmednn} & No & Yes & Yes & Yes \\
    \hline
    \cite{alani2023explainable} & No & Yes & Yes & Yes\\
    \hline
    \cite{aljuhani2023intelligent} & No & Yes & Yes & Yes \\
    \hline
    \cite{tauqeer2022cyberattacks} & No & Yes & Yes & No \\
    \hline
    \cite{Ghubaish2023} & No & Yes & Yes & No \\
    \hline
    \cite{dave2020explainable} & - & - & No & Yes \\
    \hline
    \cite{sohail2024explainable} & No & No & Yes & Yes \\
    \hline
    \cite{ashraf2024beyond} & - & - & No & Yes \\    
    \hline
    Our work & Yes & Yes & Yes & Yes \\
    \hline
\end{tabular}
\vspace{-4mm}
\end{table}

The work~\cite{HadyIDS} introduces the WUSTL-EHMS-2020 dataset, demonstrating that combining network flow metrics and biomedical data improves intrusion detection in healthcare. The authors developed an Enhanced Healthcare Monitoring System (EHMS) testbed, which leverages machine learning capabilities to address security challenges using a range of healthcare sensors. The system comprises a gateway for data collection, an IDS computer to monitor network traffic and identify anomalous activities, an attacker module to simulate real-world attack scenarios, and a server. The system enables the evaluation of how integrating network and biometric data enhances the detection of intrusions in healthcare environments.

The studies ~\cite{alani2023xmednn} and ~\cite{alani2023explainable} proposed systems to protect IoMT devices using robust classifier algorithms, leveraging the WUSTL-EHMS-2020 dataset and analyzing results with the SHAP framework. Both works provided only summary plots ranking features and focused on the impact of network features, identifying the most influential ones while noting features with minimal or no contribution to the classifier outputs. However, the studies did not explore biomedical features or emphasize the multiclass nature of the classification, leaving a significant gap in the analysis of biomedical aspects and limiting the comprehensiveness of their evaluation.

The research ~\cite{aljuhani2023intelligent} explores the balance between practicality and security in healthcare environments that utilize IoMT on a daily basis. It proposes a system that employs ensemble machine learning and deep neural networks to classify attacks and normal traffic using the WUSTL-EHMS-2020 dataset. The results are analyzed using SHAP tools, which rank the features that most significantly impact the algorithm's output. Among these, blood pressure and peripheral oxygen saturation emerge as highly influential features. Nevertheless, the study limits its analysis to feature ranking without delving deeper into the underlying reasons or implications of these findings.

The analysis by ~\cite{tauqeer2022cyberattacks} highlights the significance of IoMT and its associated security challenges, employing multiple machine learning algorithms to classify cyberattacks and normal connections using the WUSTL-EHMS-2020 dataset, achieving strong accuracy. Similarly, ~\cite{Ghubaish2023} addresses the complexities of IoMT environments and introduces a feature engineering approach for ML, evaluated on three datasets, including WUSTL-EHMS-2020. While both studies deploy algorithms for cyberattack detection, neither incorporates explainability into their methodologies.

The authors of ~\cite{sohail2024explainable} explored the use of IDS in the IoMT domain, focusing on multiclass classification of cyberattacks using three boosting algorithms. They employed the CICIoMT-2024 dataset, which contains only IoMT traffic data without biomedical data. To improve model interpretability, they integrated SHAP, offering transparency into the IDS framework's decision-making process.

Furthermore, \cite{dave2020explainable} emphasizes AI explainability for IoMT systems, demonstrating SHAP and LIME on heart disease datasets to interpret black-box models. \cite{ashraf2024beyond} similarly applies these techniques to disease prediction data based on symptoms, increasing transparency in AI decision-making.

\section{Material \& Methods} \label{sec:material}

This section describes the methodology used to analyze explainability in the WUSTL-EHMS-2020 dataset. The study applies SHAP analysis to assess the impact of network and biomedical features on intrusion detection models.

\subsection{Dataset} \label{subsec:dataset}

The WUSTL-EHMS-2020 dataset \cite{HadyIDS} was collected in a controlled testbed environment, where IoMT sensors attached to a patient's body transmitted physiological signals to a gateway device. These signals were relayed through a network switch and router to a server for real-time monitoring. Although, the transmission path was vulnerable to interception by malicious users seeking to compromise sensitive medical data. 

The dataset provides both binary and multiclass labeling. In this study, we adopt the latter, which includes three classes: benign, spoofing, and data alteration. Spoofing attacks intercept and manipulate network packets, compromising data confidentiality, while data alteration attacks modify packet contents to undermine integrity. To address these threats, an IDS was deployed in the testbed to capture both network and biomedical data, analyzing them for anomalies. The dataset consists of 16,318 samples, with 14,272 (87.5\%) labeled as benign and 2,046 (12.5\%) as attacks. It includes 45 features, comprising:

\begin{itemize}
    \item \textbf{35 network flow metrics}, such as packet transmission rates and protocol-specific metadata.
    \item \textbf{8 biometric measurements}, including heart rate and oxygen saturation.
    \item \textbf{2 classification labels}, representing the binary attack indicator (Normal or Attack) and the multiclass attack category (Normal, Spoofing, or Data Alteration).
\end{itemize}

Integrating network and biomedical features enables context-aware anomaly detection, bridging cybersecurity with human-device interactions in Healthcare 5.0. Multiclass labeling enhances classification granularity and supports SHAP-based interpretability, improving insights into intrusion detection models in healthcare environments.

\subsection{Methodology} \label{subsec:methodology}

\begin{figure}[ht]
\centering
\includegraphics[width=.75\textwidth]{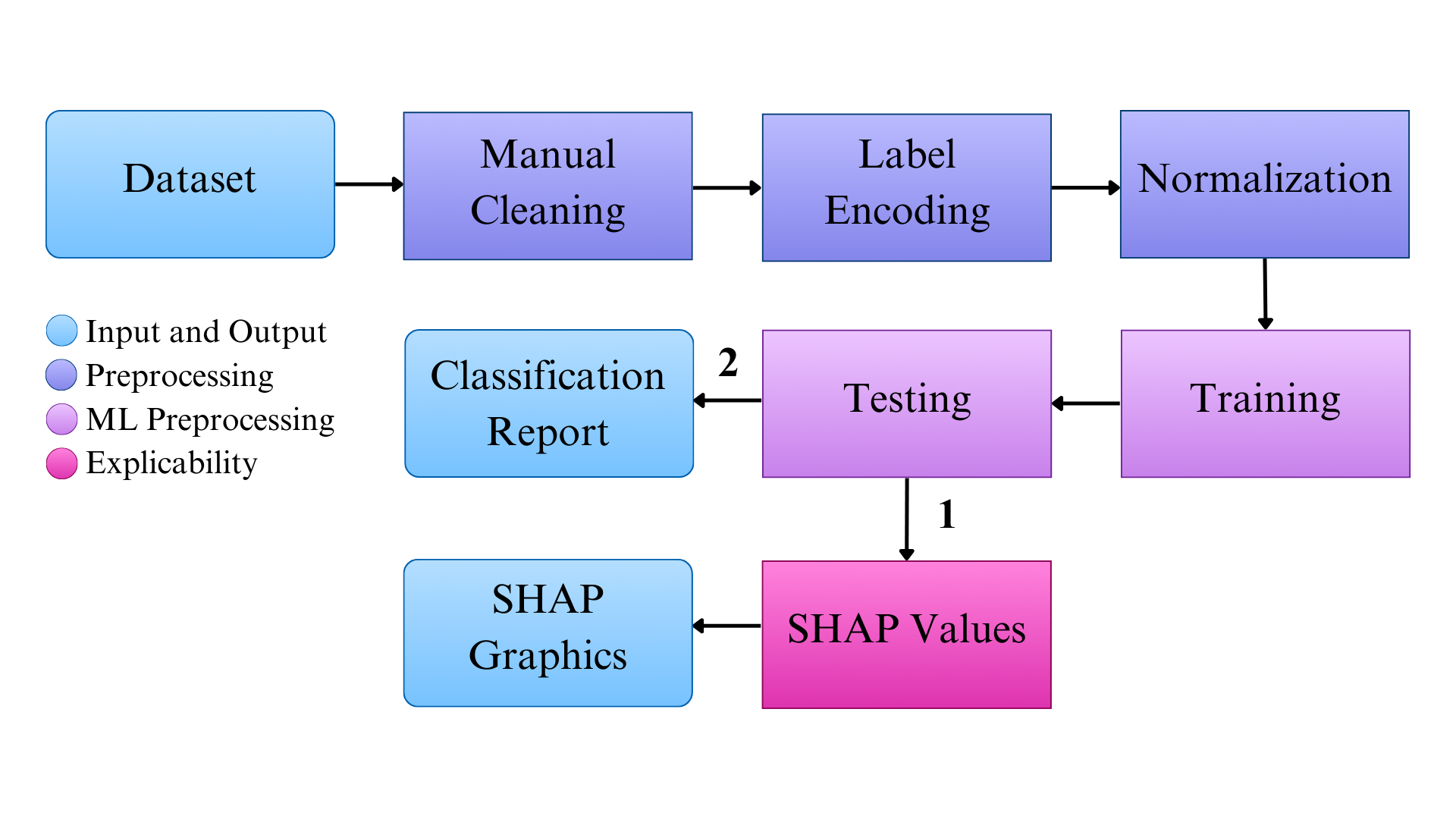}
\vspace{-7mm}
\caption{Methodology Flow Chart.}
\label{fig:flowchart}
\end{figure}

The methodology adopted to analyze the WUSTL-EHMS-2020 dataset followed a structured pipeline (Figure~\ref{fig:flowchart}), consisting of data preprocessing, model training, evaluation, and explainability.

As shown in Figure~\ref{fig:flowchart}, preprocessing was performed in multiple steps to prepare the dataset for machine learning models. Manual cleaning involved removing non-contributory features: the \texttt{Source MAC Address} and \texttt{Label} columns were discarded to prevent model overfitting, as attacks originated from a single device. The \texttt{Attack Category} was retained as the target variable, while \texttt{Dir} and \texttt{Flgs} features were dropped due to missing records. Additionally, three anomalous samples with invalid \texttt{Source Port} entries were excluded, resulting in a dataset with 16,315 samples.

Next, categorical features were encoded using scikit-learn’s LabelEncoder, ensuring compatibility with ML models. Without this transformation, features such as \texttt{Source Port} would contain NaN values, which lack SHAP interpretability. To ensure numerical stability, all features were normalized using StandardScaler. The dataset was split into 80\% training and 20\% testing subsets, as shown in Figure~\ref{fig:flowchart} (Training and Testing steps).



After preprocessing, the dataset was used to train and evaluate four classifiers: XGBoost (XGB), Random Forest (RF), Decision Tree (DT), and Support Vector Classifier (SVC). These models were selected for their ability to handle imbalanced datasets and structured tabular data. XGB was chosen for its gradient-boosting approach, which improves learning efficiency while incorporating regularization techniques to mitigate overfitting~\cite{chen2016xgboost}. RF, an ensemble of decision trees, was employed for its bagging strategy and feature randomness, enhancing generalization while maintaining feature importance interpretability. SVC was used for its ability to model nonlinear decision boundaries in high-dimensional spaces, making it effective in detecting subtle attack patterns. Finally, DT~\cite{predegosa2011scikit} was included as a simpler, rule-based classifier, offering an interpretable reference point for comparison. 

To assess classification performance, precision, recall, and F1-score were computed. Precision quantifies the proportion of correctly predicted attacks among all attack predictions, whereas recall evaluates the model's ability to identify actual attacks. Since IDSs must balance false positives and false negatives, F1-Score was selected as the harmonic mean of precision and recall, ensuring a reliable metric for assessing effectiveness.

To ensure model interpretability, SHAP was applied to assess feature importance. As indicated in Figure~\ref{fig:flowchart}, SHAP Graphics were generated to analyze how each feature influenced classification decisions. By leveraging Shapley values from game theory, SHAP provides a transparent and interpretable framework for understanding the decision-making process of intrusion detection models~\cite{lundberg2017unified}. In order to assure the results reproducibility, the materials used are publicly available\footnote{Available at: \url{https://github.com/lps5e/IA2S/}}.



\section{Results and Discussion} \label{sec:results}
This section shows a model performance evaluation. Also, it is presented and discussed the feature impact in model prediction from the XAI perspective.

\subsection{Classification Results} \label{subsec:classification}
Using the Classification Report function from the Scikit-learn\footnote{Scikit-learn library. Available at: \url{https://scikit-learn.org/}}, we computed the precision, recall, and F1-score for each label, with results presented in Table~\ref{tab:classifiers}. According to these results, all models achieved high F1-scores for Benign and Data Alteration labels. However, detecting spoofing attacks posed significant challenges, with performance varying widely across classifiers. This issue is likely influenced by the class imbalance in the dataset, where malicious samples account for only 12.5\% of the data.

\begin{table}[ht]
\centering
\caption{Testing Results With Four Classifiers}
\label{tab:classifiers}
\begin{tabular}{|c|c|c|c|c|}
    \hline
    \textbf{Model}& \textbf{Label} & \textbf{Precision} & \textbf{Recall}& \textbf{F1-Score}\\
    \hline
     & Benign & 0.98 & 0.99 & 0.99 \\
    XGB & Data Alteration & 0.99 & 0.99 & 0.99\\
     & Spoofing & 0.92 & 0.72 & 0.81\\
    \hline
     & Benign & 0.93 & 1.00 & 0.96 \\
    RF & Data Alteration & 1.00 & 0.99 & 0.99\\
     & Spoofing & 0.75 & 0.07 & 0.14\\
    \hline
     & Benign & 0.97 & 0.98 & 0.97 \\
    DT & Data Alteration & 1.00 & 1.00 & 1.00\\
     & Spoofing & 0.71 & 0.67 & 0.69 \\
    \hline
      & Benign & 0.92 & 1.00 & 0.96 \\
    SVC & Data Alteration & 0.99 & 0.99 & 0.99\\
     & Spoofing & 1.00 & 0.01 & 0.02\\
     \hline
\end{tabular}
\end{table}

\begin{figure*}[bht]
    \centering
        \subfigure[Confusion Matrix - RF.]{
            \includegraphics[width=0.43\linewidth]{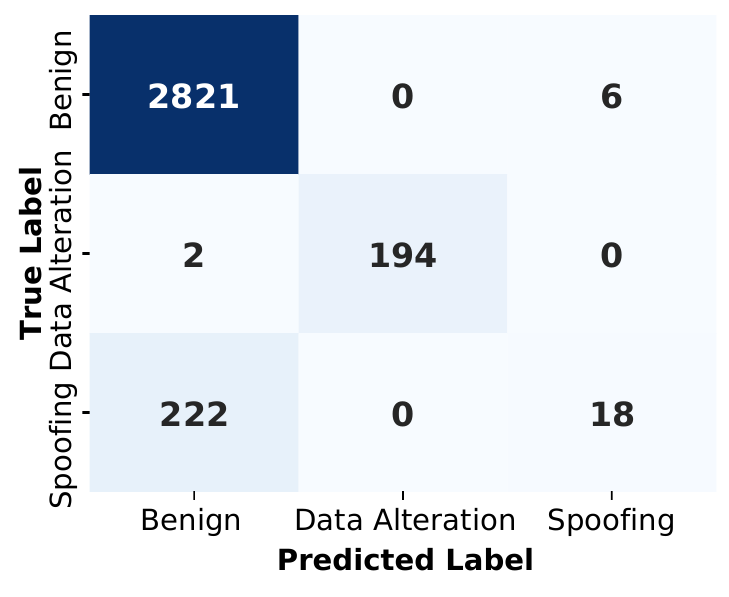}
            \label{fig:matriz-rf} 
        }
        \subfigure[Confusion Matrix - SVC.]{
            \includegraphics[width=0.43\linewidth]{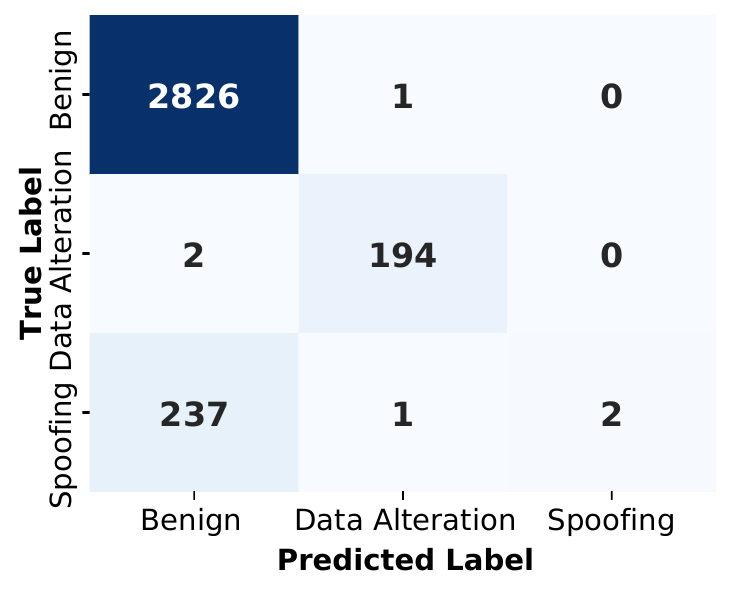}
            \label{fig:matriz-svm}
        }
\caption{Confusion matrices for RF and SVC classifiers, highlighting misclassification issues in spoofing detection.}
\label{fig:confusion-matrices}
\end{figure*}

Notably, Random Forest and Support Vector Machines performed poorly in identifying spoofing attacks, as evident in their confusion matrices (Figure~\ref{fig:confusion-matrices}). Both models exhibit a high misclassification rate, with RF misclassifying 222 spoofing samples as benign and SVC misclassifying 237 spoofing samples as benign, leading to very low recall scores (Table~\ref{tab:classifiers}). This suggests that these models struggle to distinguish spoofing from benign traffic, potentially due to overlapping feature distributions. Deeper analysis suggests the RF’s tendency to favor the class with more number of samples---imbalance datasets results in weak learning of the spoofing class, whereas SVC’s margin-maximization criterion and choice of kernel may fail to capture the nuanced patterns of spoofed samples, compounding their misclassification rates.

In contrast, XGBoost demonstrated more consistent performance across all labels, particularly in spoofing detection, achieving a recall score of 0.72, significantly outperforming RF (0.07) and SVC (0.01). XGB’s superior performance can be attributed to its gradient-boosting framework, which effectively handles imbalanced datasets and reduces overfitting. Consequently, XGB was selected for SHAP-based explainability analysis, as it provides the most reliable attack classification while maintaining model interpretability.


\subsection{Explaining Feature Contributions using SHAP Analysis} \label{subsec:shap}
This section presents SHAP explainability plots to identify the most relevant features in predicting each attack category. Two plot types were generated: (i) the summary bar plot, which ranks features by their average SHAP values, and (ii) the summary plot, which provides a detailed view of feature contributions across samples.



\begin{figure}[ht]
\centering
\includegraphics[width=.7\textwidth]{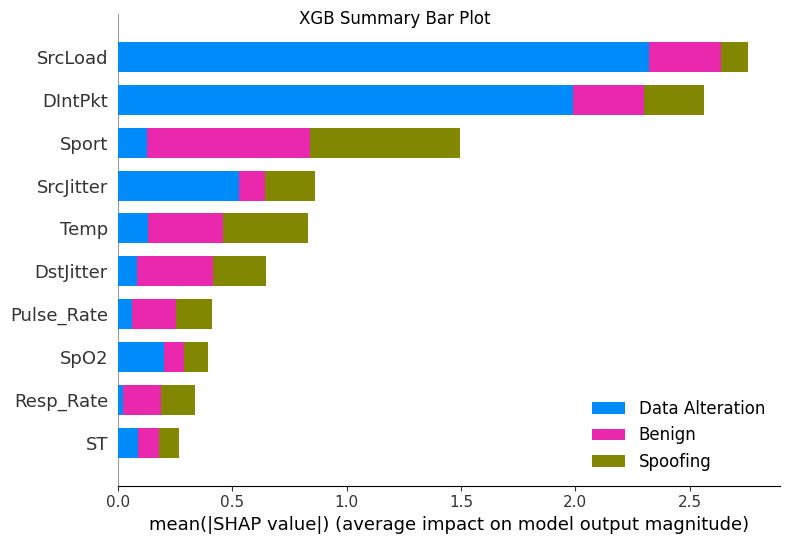}
\caption{SHAP Summary Bar Plot.}
\label{fig:xgbbar}
\vspace{-6mm}
\end{figure}

Figure~\ref{fig:xgbbar} illustrates the SHAP summary bar plot, ranking the ten most influential features by their mean SHAP values. As expected, Source Load (\texttt{SrcLoad}) appears as the most impactful feature in Data Alteration detection, reinforcing the idea that attackers often manipulate traffic rates to evade anomaly detection. Attackers may introduce artificial delays, reduce transmission rates, or flood the network with modified packets, making \texttt{SrcLoad} variations a strong indicator of malicious activity. Similarly, Destination Interpacket Arrival Time (\texttt{DIntPkt}) emerges as another critical feature across all classifications. Higher interpacket arrival times can suggest stealthy attack techniques, where adversaries intentionally slow down data transmission to avoid triggering rate-based IDS mechanisms. The Source Port (\texttt{Sport}) also plays a major role, particularly in Benign and Spoofing classifications, aligning with well-documented spoofing strategies where attackers modify source port numbers to masquerade as legitimate traffic~\cite{tinshu2024comprehensive}. The presence of \texttt{DstJitter} and \texttt{SrcJitter} in the top-ranked features further supports the hypothesis that attackers tend to introduce irregular packet timing, a common evasion technique to avoid signature-based IDS detection.


Interestingly, several biomedical features also appear among the most relevant indicators of intrusions, particularly in spoofing detection. Features such as Temperature (\texttt{Temp}), Pulse Rate, Peripheral Oxygen Saturation (\texttt{SpO2}), Respiratory Rate (\texttt{Resp Rate}), and ECG ST Segment (\texttt{ST}) demonstrate significant contributions to classification, reinforcing the argument that physiological signals should not be overlooked in cybersecurity applications. One possible explanation is that spoofing attacks may introduce anomalies in biomedical data streams, either through sensor interference, adversarial data injections, or inconsistencies in physiological responses during an attack event. For instance, anomalous fluctuations in temperature or heart rate could indicate tampering with wearable or implantable medical devices, as seen in previous research on biomedical sensor vulnerabilities~\cite{khan2020biometric}. These findings emphasize the potential for hybrid IDS models, where integrating both network and biomedical features can enhance the detection of cyber-physical attacks in Healthcare 5.0 environments.

\begin{figure*}[bht]
    \centering
        \subfigure[SHAP Summary Plot Benign.]{
            \includegraphics[width=0.482\linewidth]{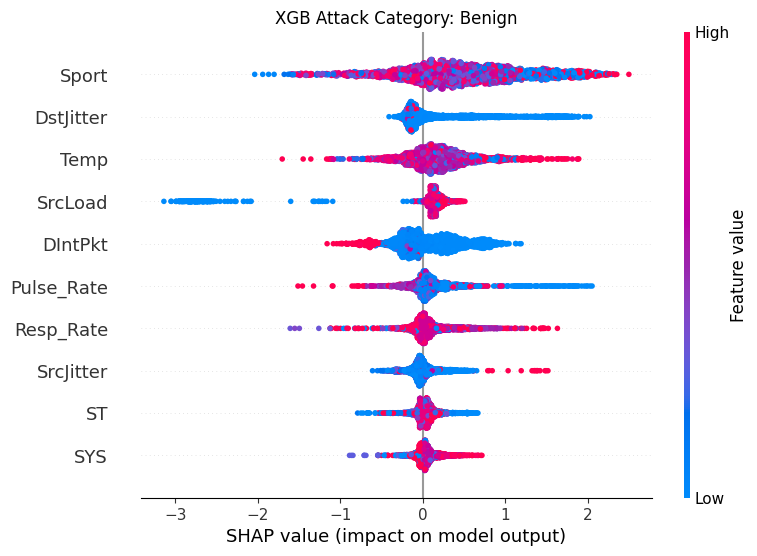}
            \label{fig:xgbsummarybenign}
        }
        \subfigure[SHAP Summary Plot Data Alteration.]{
            \includegraphics[width=0.482\linewidth]{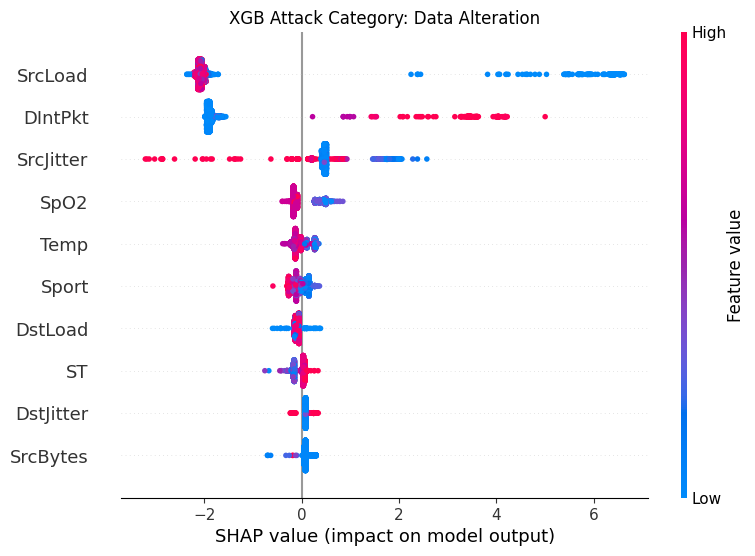}
            \label{fig:xgbsummarydata} 
        }
        \hfill
        \subfigure[SHAP Summary Spoofing.]{
            \includegraphics[width=0.482\linewidth]{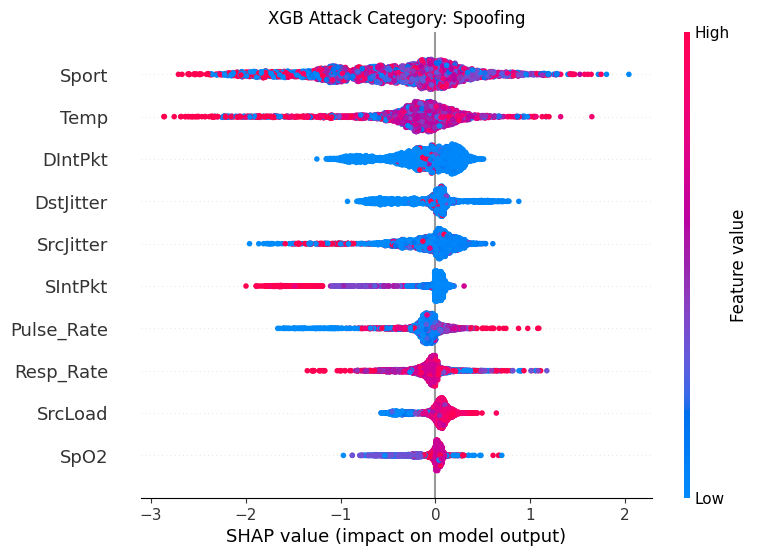}
            \label{fig:xgbsummaryspoofing}
        }
\caption{SHAP plots showing feature importance for (a) benign, (b) data alteration and (c) spoofing attacks.}
\label{fig:subcenariosAeB}
\vspace{-4mm}
\end{figure*}

Sequentially, using SHAP summary plots, each classification label will be thoroughly examined for a more detailed approach. Figure \ref{fig:xgbsummarybenign} presents the SHAP summary plot for the Benign label, highlighting the ten most relevant features. Positive SHAP values indicate a higher likelihood of benign classification, while negative values push the prediction toward a non-benign label. The color scale represents the magnitude of each feature value, from low (blue) to high (red).

The most relevant feature is \texttt{Sport}. Lower port numbers (blue, left-skewed dots) correlate with non-benign traffic, whereas higher and medium port numbers are linked to benign traffic. This divergence likely stems from differences in port allocation mechanisms, as benign traffic typically adheres to standard automated port assignment protocols, whereas malicious traffic frequently employs manually specified ports. Other network-related features are also present, \texttt{DstJitter} and \texttt{DIntPkt} indicate that lower variation in time between packets and lower interpacket arrival time in destination contribute to identifying a sample as Benign. However, the opposite occurs for \texttt{SrcJitter} and \texttt{SrcLoad}, where lower variation and lower load at the source push the classification as a non-benign classification. Additionally, biomedical features such as \texttt{Temp}, \texttt{Pulse\_Rate}, \texttt{Resp\_Rate}, \texttt{ST}, and \texttt{SYS} also contribute to the classifier's decision-making. Their presence among the most relevant features suggests that physiological data might hold useful information for distinguishing benign traffic.


Figure \ref{fig:xgbsummarydata} represents the SHAP Plot for the Data Alteration label, which refers to instances of packet modification that violate data integrity. \texttt{SrcLoad} is the attribute with the strongest predictive influence. According to the analysis, smaller \texttt{SrcLoad} values are highly correlated with classifications that lean toward malicious, while higher values cause predictions to lean toward normal traffic. On the other hand, \texttt{DIntPkt} shows that high interpacket arrival times are relevant for detecting Data Alteration attacks. This pattern is consistent with well-known evasion strategies used in cyberattacks, where malicious actors frequently reduce data transfer speeds to evade detection systems, resulting in low bit rates and high arrival times. Additionally, \texttt{SrcJitter} with high values contributes to reducing SHAP values, indicating a lower occurrence of Data Alteration attacks when those values are high. Furthermore, we observe that biomedical features (\texttt{SpO2}, \texttt{Temp}, \texttt{ST}) appear in the plot with SHAP values fluctuating within a limited range, suggesting that while they have some level of variability, their impact on the model’s decision-making process is less pronounced compared to network-related attributes. This indicates that, for Data Alteration attacks, the model relies primarily on network-based features, though the potential role of biomedical data in other attack scenarios should not be disregarded.

Lastly, Figure \ref{fig:xgbsummaryspoofing} highlights the most influential features for detecting spoofing attacks. The network feature \texttt{Sport} and the biomedical feature \texttt{Temp} exhibit the strongest impact on model predictions. Lower \texttt{Sport} values (blue) tend to reduce the likelihood of a spoofing attack, whereas higher values (red) push the model toward predicting spoofing. Similarly, \texttt{Temp} exhibits a wide range of SHAP values, indicating that variations in body temperature are considered by the model when predicting spoofing attempts, though the exact nature of this relationship requires further analysis.

Network-related attributes such as \texttt{DIntPkt}, \texttt{DstJitter}, and \texttt{SrcJitter} suggest that lower values increase the likelihood of an attack, reinforcing the idea that spoofing attacks often involve altered timing characteristics. Additionally, \texttt{SrcLoad} suggests that increased source load may be indicative of spoofing activity, possibly due to altered traffic patterns in such attacks.

Overall, spoofing detection appears weaker compared to other attack categories, as previously discussed in Section \ref{subsec:classification} and summarized in Table \ref{tab:classifiers}. This limitation may stem from the nature of spoofing attacks.


\section{Conclusion and Future Works} \label{sec:conclusao}

Existing datasets and experimental frameworks in IDS predominantly focus on network traffic data, with empirical analyses largely confined to network-layer attacks. In IoT-health ecosystems, adversarial manipulations of biomedical sensor data remain critically understudied due to this methodological gap. By incorporating cutting-edge technologies, the Healthcare 5.0 idea seeks to address this problem. Through SHAP value analysis, our investigation reveals that network traffic features dominate categorization outcomes in all scenarios. In contrast, biomedical sensor inputs—such as heart rate, blood pressure, and glucose monitoring data—demonstrate relevance in most cases. However, their low attribution scores in data alteration attack detection indicate negligible influence on model decisions. For future work, we intend to (i) investigate the inclusion of more intricate datasets, particularly those incorporating clinically significant biomedical attack vectors, addressing the need for evaluation in diverse clinical contexts, and (ii) perform XAI analysis across biomedical, network, and combined data, exploring three different groups.

\section*{Acknowledgments}
This research effort is sponsored in part by resources from ``Edital PRPGP/UFSM N.50/2024 - Programa de Fortalecimento e Redução de Assimetrias da Pós-Graduação da UFSM''. Also, this work is partially supported by CIARS RITEs/FAPERGS project.

\footnotesize
\bibliographystyle{sbc}
\bibliography{main}

\end{document}